\documentclass[notitlepage,superscriptaddress,aps,prd,nofootinbib,twocolumn,showpacs]{revtex4-1}
\usepackage[british,UKenglish,USenglish,american]{babel}
\usepackage{amsfonts}
\usepackage{amsmath}
\usepackage{amssymb}
\usepackage{graphicx}
\usepackage{mathrsfs}
\usepackage{hhline,color}
\usepackage{pifont,ulem}
\usepackage{lmodern}
\usepackage{epsfig}
\usepackage{enumitem} 
\usepackage{ulem}
\usepackage[utf8]{inputenc}
\usepackage{multirow}

\usepackage{braket}

\def\beq{\begin{equation}}
\def\eeq{\end{equation}}
\def\beqa{\begin{eqnarray}}
\def\eeqa{\end{eqnarray}}
\def\ban{\begin{eqnarray*}}
\def\ean{\end{eqnarray*}}
\def\bi{\begin{itemize}}
\def\ei{\end{itemize}}

\usepackage{physics}

\usepackage{slashed}

\usepackage{mathrsfs}
\usepackage{amssymb}
\usepackage{bbm}

\usepackage{nicefrac}

\usepackage{microtype}

\newcommand\numberthis{\addtocounter{equation}{1}\tag{\theequation}}

\begin{document}

\title{
Role of the conserved charges in the chiral symmetry restoration phase
transition 
}

\author{Pedro Costa} 
\email{pcosta@uc.pt}
\affiliation{CFisUC, Department of Physics,
University of Coimbra, P-3004 - 516  Coimbra, Portugal}
\author{Renan Câmara Pereira}
\email{renan.pereira@student.uc.pt}
\affiliation{CFisUC, Department of Physics,
University of Coimbra, P-3004 - 516  Coimbra, Portugal}
\author{Constan\c ca Provid\^encia}
\email{cp@fis.uc.pt}
\affiliation{CFisUC, Department of Physics,
University of Coimbra, P-3004 - 516  Coimbra, Portugal}

\date{\today}

\begin{abstract}
The effect of conserved baryon, isospin and strangeness charges on the
behavior of phase transitions in dense matter is studied. Baryonic
matter is described within the three-flavor
Polyakov$-$Nambu$-$Jona-Lasinio model and several  charge fractions
$Y_Q$ are considered. The role of the  vector interaction, which can be
important to describe dense systems, is discussed. Special attention is given 
to the case with charge fraction $Y_Q=0.4$, due to its importance in heavy-ion 
collisions and core-collapse supernova matter. It is shown that the possible 
formation of chiral-symmetric quark matter in the laboratory will be favored in 
asymmetric  matter. Besides,  the inclusion of the vector interaction reinforces 
the formation of quark matter at lower densities.

\end{abstract}


\maketitle

\section{Introduction}

We investigate the phase transition  associated with the restoration of
chiral symmetry in a system with more than one conserved charge
(multicomponent systems): baryonic charge, isospin, and
strangeness. This is relevant for scenarios where asymmetric matter
occurs, like in heavy ion collisions (HICs) \cite{Greiner:1987tg} and
compact stars \cite{Glendenning:1992vb}. Given a fixed electric-to-baryon 
charge ratio, $Q/B\neq0.5$, this additional
isospin degree of freedom cannot be exploited by the system in pure
phases. In the mixed phase, however, the total asymmetry is constant,
while the local asymmetries of each phase can be different. For more
than one globally conserved charge, such as baryon, isospin, and/or
strangeness, phase equilibrium has to be implemented by imposing Gibbs
rules, which modify both the structure of the mixed phase and the
determination of the transition point. When only one globally
conserved charge is allowed, the phase equilibrium is obtained by
a Maxwell construction at constant pressure.

In the modelling of a first-order phase transition with density as the
order parameter, it is often assumed that the coexistence region can
be obtained by a  Maxwell construction at constant pressure.
However, this is not true if different charges correspond to good
quantum numbers, and in that case full Gibbs conditions must be
applied \cite{Glendenning:1997wn}. In a multicomponent system,
the local concentrations of charges vary during the crossing of a
phase-coexistence region,  as well as the  pressure and
baryonic chemical potential.
Considering the phase diagram in the temperature-pressure plane, for a given
temperature, the transition occurs over a range of pressures.
This effect is well known in plasma and condensed matter physics and was recently
applied to the hadronic sector  in Ref. \cite{Hempel:2013tfa}.

In Ref. \cite{Muller:1995ji} the liquid gas phase transition in asymmetric 
nuclear matter, a system with more than one conserved charge, was intensely 
discussed. In fact, when building the equation of state for a neutron star, both
electric and baryonic charges have to be conserved. The description of
the phase transition from hadrons to quarks inside such an object is obtained by
applying a Gibbs construction in order to identify   the limits of the mixed 
phase \cite{Glendenning:1992vb}.
Finding the equilibrium points that satisfy all Gibbs conditions may be 
cumbersome, but in Ref. \cite{Ducoin:2005aa} a new statistical method was 
introduced to study the thermodynamics of a multifluid system that reduced the 
problem to Maxwell constructions. 
It consists of keeping only one density fixed and replacing the others by 
their intensive conjugated variables.

We will perform our study within the (2+1)-flavor Polyakov-loop-extended Nambu$-$Jona-Lasinio (PNJL) model
\cite{Fu:2007xc,Ciminale:2007sr,Fukushima:2008wg,Abuki:2008nm,Costa:2010zw}. 
Models of the Nambu$-$Jona-Lasinio type are effective field theories
describing the basic mechanisms that drive the spontaneous breaking of
chiral symmetry \cite{Asakawa:1989bq} a key feature of quantum chromodynamics 
(QCD), and are widely used to study the phase diagram of strongly-interacting
matter (also called the QCD phase diagram), hadron phenomenology, and the quark phase 
of the neutron star equation of state. The inclusion of the Polyakov loop in
these models allows to better reproduce lattice-QCD results and to
study the confinement-deconfinement transition. We consider a  model
with both a vector-isoscalar and a vector-isovector contribution, with
equal coupling constants. Starting from a QCD-inspired color current–current interaction, 
these vector channels and the tensor channel can be related to the 
scalar-pseudoscalar channel using a Fierz transformation into color-singlet 
channels \cite{Bratovic:2012qs}.  
Indeed, in \cite{Braun:2018bik,Braun:2017srn}, a so-called Fierz-complete minimal 
set of channels was proposed from which any other interaction channel can be 
derived using Fierz transformations. In such an approach,the 
scalar-pseudoscalar and vector couplings are not independent and indeed they 
can be fixed by the scalar-pseudoscalar coupling \cite{Lutz:1992dv,Bratovic:2012qs}. 
From a phenomenological point of view, the vector coupling can also be fixed in 
the vacuum by requiring the model to be able to reproduce the vacuum masses of 
vector mesons. However, the vector interactions are known to couple to density 
degrees of freedom and the overall magnitude of the vector interaction might be 
density dependent, i.e., chemical potential dependent. 
In fact, there is still no constraint for the choice of an induced 
vector coupling at finite density. For example, we do not know if vector 
interactions induce a more attractive or repulsive interaction (and thus its sign is 
also unknown). 
Several studies of the QCD phase diagram and the neutron star equation of state 
have considered the vector couplings as free parameters (see, e.g.,
\cite{Hanauske:2001nc,Fukushima:2008wg,Bonanno:2011ch,Logoteta:2013ipa,Beisitzer:2014kea,Zacchi:2015lwa,Costa:2016vbb,Pereira:2016dfg}):
\begin{itemize}
	\item[-] It has been shown that the vector interaction has a strong influence on 
  the chiral-symmetry-restoration phase transition, making (for a strong enough
  interaction) a first order phase transition turns into a smooth crossover.
  \item[-] The vector channel is important for modelling the quark degrees of freedom 
  inside the core of a neutron star \cite{Pereira:2016dfg}.
\end{itemize}

In Refs. \cite{Sasaki:2007db,Sasaki:2007qh} the authors have shown that 
divergent density fluctuations result from spinodal decomposition in a 
nonequilibrium first-order chiral phase transition; in particular, the specific 
heat and charge susceptibilities diverge at the isothermal spinodal
lines.  The study was performed within the NJL model but the same
conclusions are expected to be  generally true.  Understanding how the
charge asymmetry may affect the  metastable and unstable regions of
the phase diagram, where this divergent behavior  is  also
expected, is one of the objectives of our work.

This paper is organized as follows. In Sec. \ref{sec:model} the PNJL model, 
which will be used to model dense baryonic matter, is introduced. 
In Sec. \ref{sec:Results}, the results of considering more then one conserved 
charge in the PNJL model are presented and the phase transitions for different 
scenarios are shown. Finally, in Sec. \ref{sec:conclusions} we present out 
conclusions and describe some perspectives for future work.

\section{Model and formalism}
\label{sec:model}

The (2+1)-flavor Lagrangian density for the PNJL model reads
\begin{align*}
\mathcal{L} & =
\bar{\psi} 
\qty(i\slashed{D} - \hat{m} + \hat{\mu} \gamma^0 
)
\psi 
\\
& + G_S  \sum_{a=0}^8
\qty[ \qty(\bar{\psi} \lambda^a \psi)^2 + 
\qty(\bar{\psi} i \gamma^5 \lambda^a \psi)^2 ]
\\
& - G_D 
\qty(
\det\qty[ \bar{\psi} \qty(1 + \gamma_5) \psi ] +
\det\qty[ \bar{\psi} \qty(1 - \gamma_5) \psi ] 
)
\\
& - G_V \sum_{a=0}^8 
\qty[ 
(\bar{\psi} \gamma^\mu\lambda^a \psi)^2 +  
(\bar{\psi} \gamma^\mu\gamma_5\lambda^a \psi)^2 
]
\\
& - \mathcal{U} \qty(\Phi,\bar{\Phi}; T ) 
.
\numberthis
\label{eq:SU3_NJL_lagrangian}
\end{align*} 
Here the quark field is represented by $\psi = (u,d,s)^T$ in flavor space, and 
${\hat m}= {\rm diag}_f (m_u,m_d,m_s)$ is the corresponding (current) diagonal 
mass matrix. Finite-density effects are included by considering a finite quark 
chemical potential matrix, $\hat{\mu}= {\rm diag}_f \qty(\mu_u,\mu_d,\mu_s)$. 
The Lagrangian includes a scalar-pseudoscalar interaction which 
spontaneously breaks chiral symmetry in the vacuum by generating a quark-antiquark 
condensate. Also present, is the so-called Kobayashi-Maskawa-'t Hooft 
interaction, responsible for the generation of six-fermion interactions
\cite{Klevansky:1992qe,Hatsuda:1994pi} which explicitly break the
$U_A(1)$ symmetry and correctly reproduce the observed hadron spectra. 
We also include in the model a vector interaction. This term includes both a 
vector-isoscalar and a vector-isovector interaction with a coupling constant 
$G_V$ \cite{Mishustin:2000ss}.

Since we are not interested in fixing the value of this parameter by reproducing
vector-meson masses, we will study the effect of having a finite ratio
$\zeta=G_V/G_S$ which is known to be important in the study of neutron
stars \cite{Bonanno:2011ch,Pereira:2016dfg} or the QCD phase diagram 
\cite{Ferreira:2018sun}. Indeed, even fixing the vector interaction in the 
vacuum does not restrict its possible in-medium dependence since it couples to 
density degrees of freedom.

The PNJL model is also capable of describing the statistical 
confinement-deconfinement transition, with the breaking of $Z(N_c)$ symmetry. 
The quark fields are minimally coupled to a background gluonic field in the 
temporal direction, $A_4^0$, through the covariant derivative 
$D_\mu = \partial_\mu - A_4^0 \delta_\mu^0$. 
An approximate order parameter for this transition is the Polyakov loop $\Phi$. 
In the confined phase $\Phi \to 0$, while in the deconfined phase 
$\Phi \to 1$.  In this model, the effective potential 
$\mathcal{U}\qty(\Phi,\bar{\Phi};T )$ is built using the Ginzburg-Landau theory 
of phase transitions: at low temperatures the $Z\qty(N_c)$ symmetry holds, while 
at high temperatures, it is broken. We choose to adopt the effective potential 
proposed in Refs. \cite{Fukushima:2003fw,Ratti:2005jh,Roessner:2006xn}:
\begin{align*}
\frac{\mathcal{U}\qty(\Phi,\bar{\Phi};T)}{T^4} = 
&-\frac{1}{2} a\qty( T ) \bar{\Phi} \Phi 
\\
& + b\qty( T ) 
\ln 
\qty[ 
1-6\Phi\bar{\Phi}+4(\Phi^3+\bar{\Phi}^3)-3(\Phi\bar{\Phi})^2
],
\numberthis
\label{eq:Polyakov.loop.potential}
\end{align*} 
with the $T$-dependent parameters \cite{Roessner:2006xn}
\begin{align*}
a\qty( T ) = a_0 + a_1 \qty( \frac{T_0}{T} ) + a_2 \qty( \frac{T_0}{T} )^2 
\quad , \quad
b\qty( T ) =  b_3 \qty( \frac{T_0}{T} )^3 .
\end{align*}
Its parametrization values are $a_0 = 3.51$, $a_1 = -2.47$, $a_2 = 15.2$, 
and $b_3 = -1.75$ \cite{Roessner:2006xn} obtained with $T_0=270$ MeV to 
reproduce the lattice QCD result. Due to the presence of quarks we  will rescale the 
critical temperature to $T_0=210$ MeV. 

The PNJL model is nonrenormalizable and a regularization scheme has to be 
introduced to deal with nonconvergent integrals in the model. 
In the present work, the divergent ultraviolet sea-quark integrals are 
regularized by a sharp cutoff $\Lambda$ in three-momentum space.
For the NJL model parametrization, we consider:
$\Lambda = 602.3$ MeV, $m_u= m_d=5.5$ MeV, $m_s=140.7$ MeV, 
$G_S \Lambda^2= 1.835$, and $G_D \Lambda^5=12.36$ \cite{Rehberg:1995kh}.

Our goal is to describe a multicomponent system in which the local 
concentrations of charges are kept fixed across a phase-coexistence region.
Although we are discussing the phase transition within a three-component system
composed of quarks $u$, $d$ and $s$, the results we present are defined by the 
first two alone. The onset of $s$ quarks occurs for much larger baryonic 
densities and chemical potentials than the ones shown in the 
following figures. 
This was discussed in Ref. \cite{Ferreira:2018sun} and it was shown that the 
$s$ quark affects the QCD phase diagram only for a baryonic chemical potential 
$\mu_B\gtrsim 1400$ MeV. 

Since we will be discussing the behavior of matter with a fixed charge
fraction with zero strangeness density ($\rho_s/\rho_B=Y_S=0$), 
the Gibbs free energy, $G$, is given by \cite{Hempel:2013tfa}:
\begin{align}
G=B \mu_B+Q\mu_Q= B \mu_{BQ} ,
\label{gibbs_free_energy_new_def}
\end{align}
where $B$ and $Q$ are the baryon and charge number, respectively, and
$\mu_{BQ}=\mu_B+Y_Q\mu_Q$ is the chemical potential which is kept constant during 
a phase transition at constant charge fraction \cite{Hempel:2013tfa}, 
if the strangeness charge is zero. 
As shown in Ref. \cite{Hempel:2013tfa}, the following interphase chemical equilibrium 
condition is imposed:
$\mu_{BQ}^I =\mu_{BQ}^{II}$, with the local Gibbs free energy per baryon 
$\mu^i_{BQ}=\mu^i_B+Y^i_Q\mu^i_Q,$ where $i=I$, and $II$ designates the phase with broken chiral symmetry and the chiral-symmetric phase, respectively.

\section{Results}
\label{sec:Results}

In the following, we discuss the first-order phase transition associated with 
the restoration  of chiral symmetry in asymmetric  quark matter, both with and without vector interactions. 
This will be implemented by fixing the value of $Y_Q$ and by the conservation 
of baryon number $B$ and total net strangeness $S = 0$. 
Due to its relevance in HICs, namely, for Au-Au or Pb-Pb collisions, we will 
frequently choose the charge fraction $Y_Q = 0.4$ and $Y_S = 0$ 
\cite{Sissakian:2006dn}. 
Another interesting scenario  for  asymmetric quark matter is the one occurring 
in core-collapse supernova matter  where matter has a proton fraction below 0.4. 
Scenarios with isospin charge conservation in strong interaction were
also studied in Refs. \cite{Cavagnoli:2010yb,Shao:2011fk,Shao:2012tu}.
Although, strangeness may occur due to the weak interaction, in the following 
discussion we consider $Y_S=0$.

We first analyze the phase diagram of the PNJL model with $Y_Q=0.5$. 
This specific value for $Y_Q$ corresponds to  describing a one-fluid matter 
consisting of symmetric matter, with equal amount of quarks $u$ and $d$ 
($\rho_u = \rho_d$, $\rho_s = 0$), and, in this case, the phase transition is
governed by $\mu_B$.  The same concerns the study of pure neutron
matter with charge fraction $Y_Q=0$, which we will also refer to in the
following.
To study the effect of a finite repulsive vector interaction, we will
present results for models with $\zeta=G_V/G_S=0$ and $\zeta=G_V/G_S=0.5$. 
The last value was chosen in order to have the critical end point (CEP) at 
$T\approx0$ for $Y_Q = 0$.

\begin{figure*}[t!]
	\centering
	\includegraphics[width=0.49\linewidth]{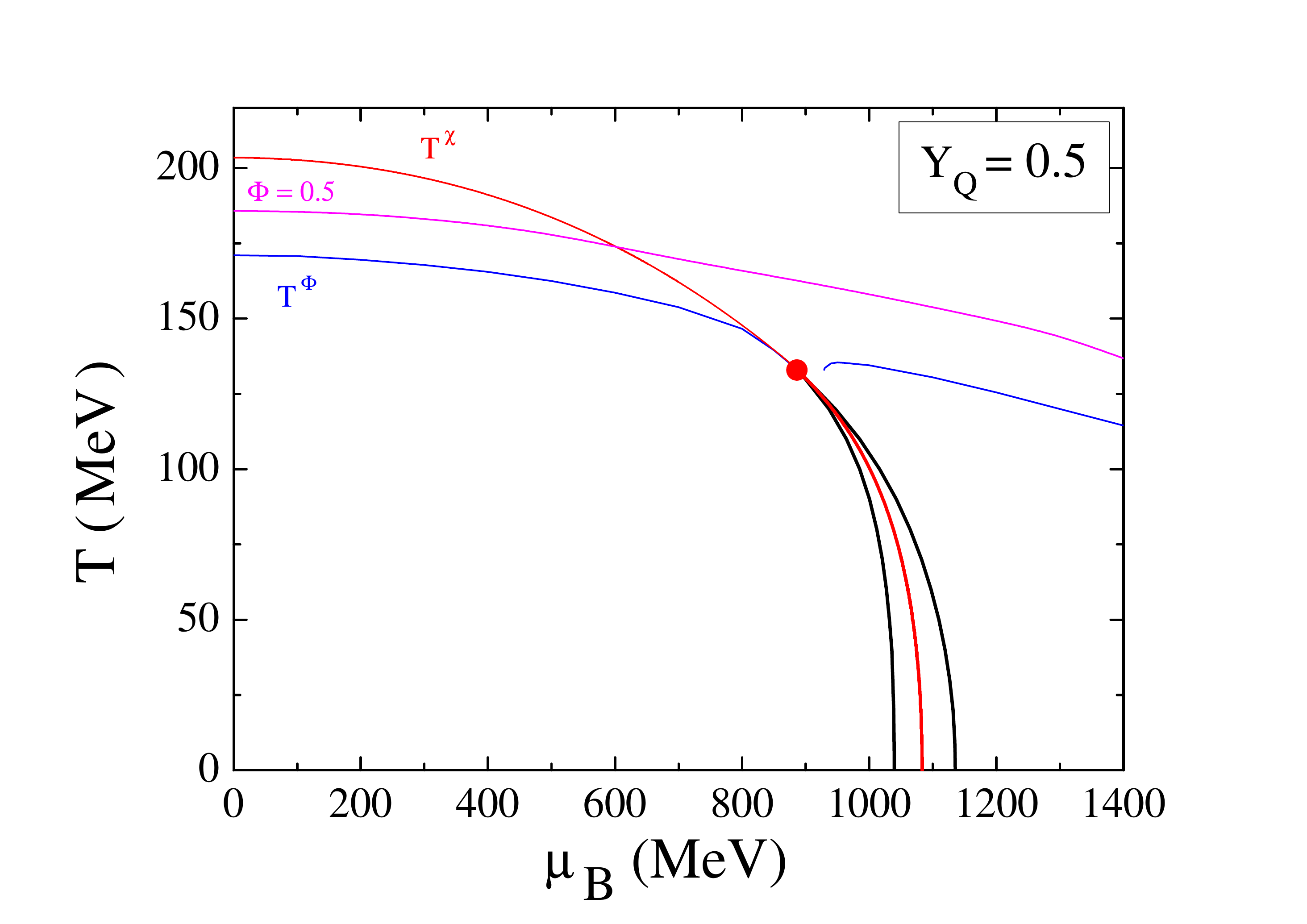}\hspace{-0.9cm}
	\includegraphics[width=0.49\linewidth]{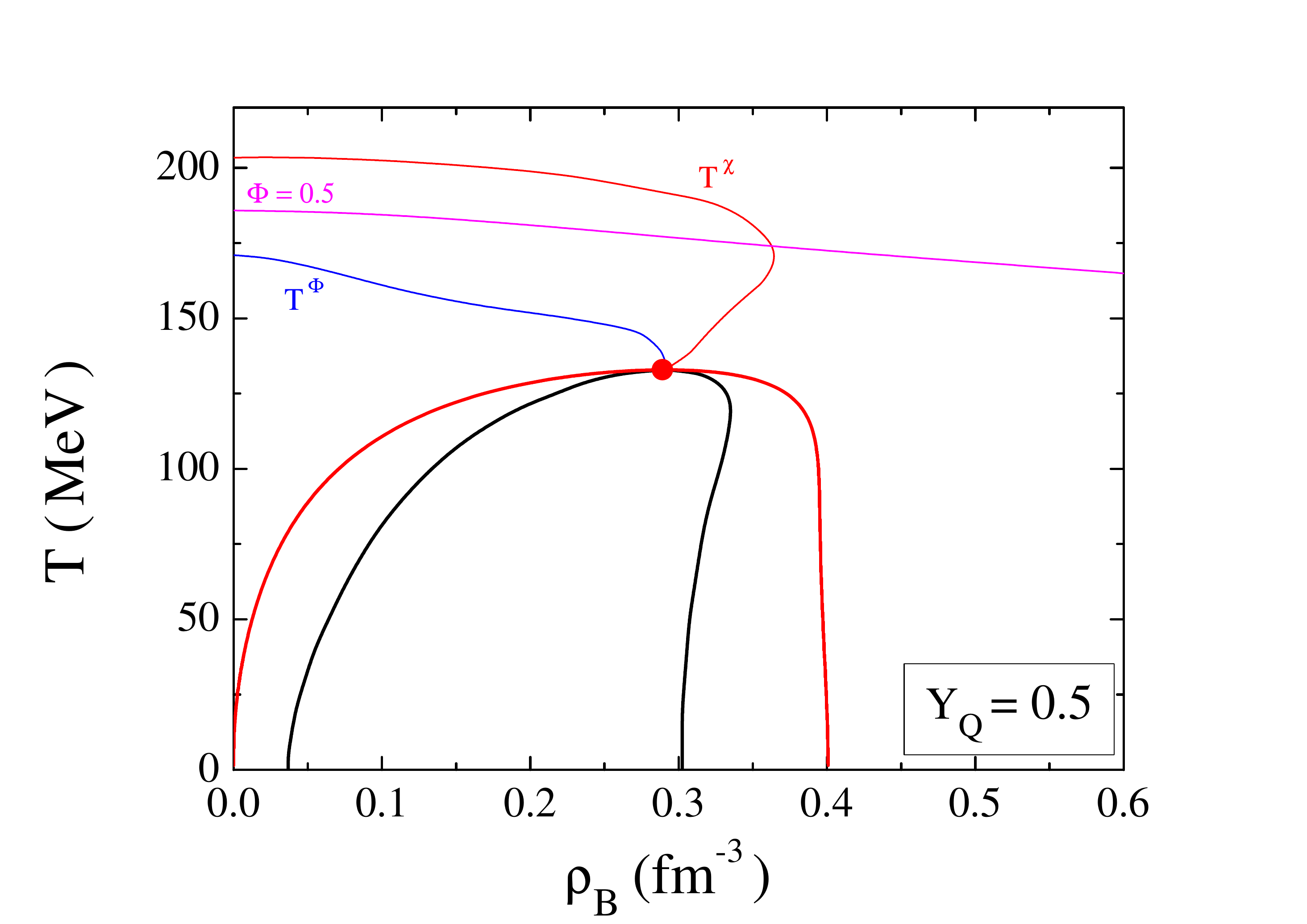}
	\caption{Phase diagram of the PNJL model with $Y_Q=0.5$, in the $T-\mu_B$ 
          (left) and $T-\rho_B$ (right) planes for the model with 
          $T_0= 210$ MeV.  
          The red lines are the crossover (thin lines) and first-order chiral 
          transition (thick lines), the blue lines define the deconfinement 
          crossover and the purple lines represent when $\Phi=0.5$. 
          The black lines are the spinodal lines and the red dot is the CEP in this scenario.
          }
	\label{fig:01}
\end{figure*}

In Fig. \ref{fig:01} the phase diagram of the model is shown in terms of the  
baryonic chemical potential (left panel) and the  baryonic density (right panel). 
In both panels the spinodal, binodal, and crossover  lines are represented by black, thick red, and thin red lines, respectively, and the deconfinement 
crossover is represented by the  blue lines. 
In the right panel, the regions between the binodal and spinodal lines and
the region inside the spinodal section correspond to
regions of metastable and unstable matter, respectively. They are reached only
during a  nonequilibrium evolution of the system. 
Under the present conditions, homogeneous chiral-symmetric quark matter is 
attained at densities above 0.4 fm$^{-3}$. However, below these densities 
chiral-symmetric quark matter will have the form of clusterized matter. 
One of the main objectives of the present study is to understand under which 
conditions chiral-symmetric quark matter is more favorably formed.

\begin{figure*}[t!]
	\centering
	\includegraphics[width=0.49\linewidth]{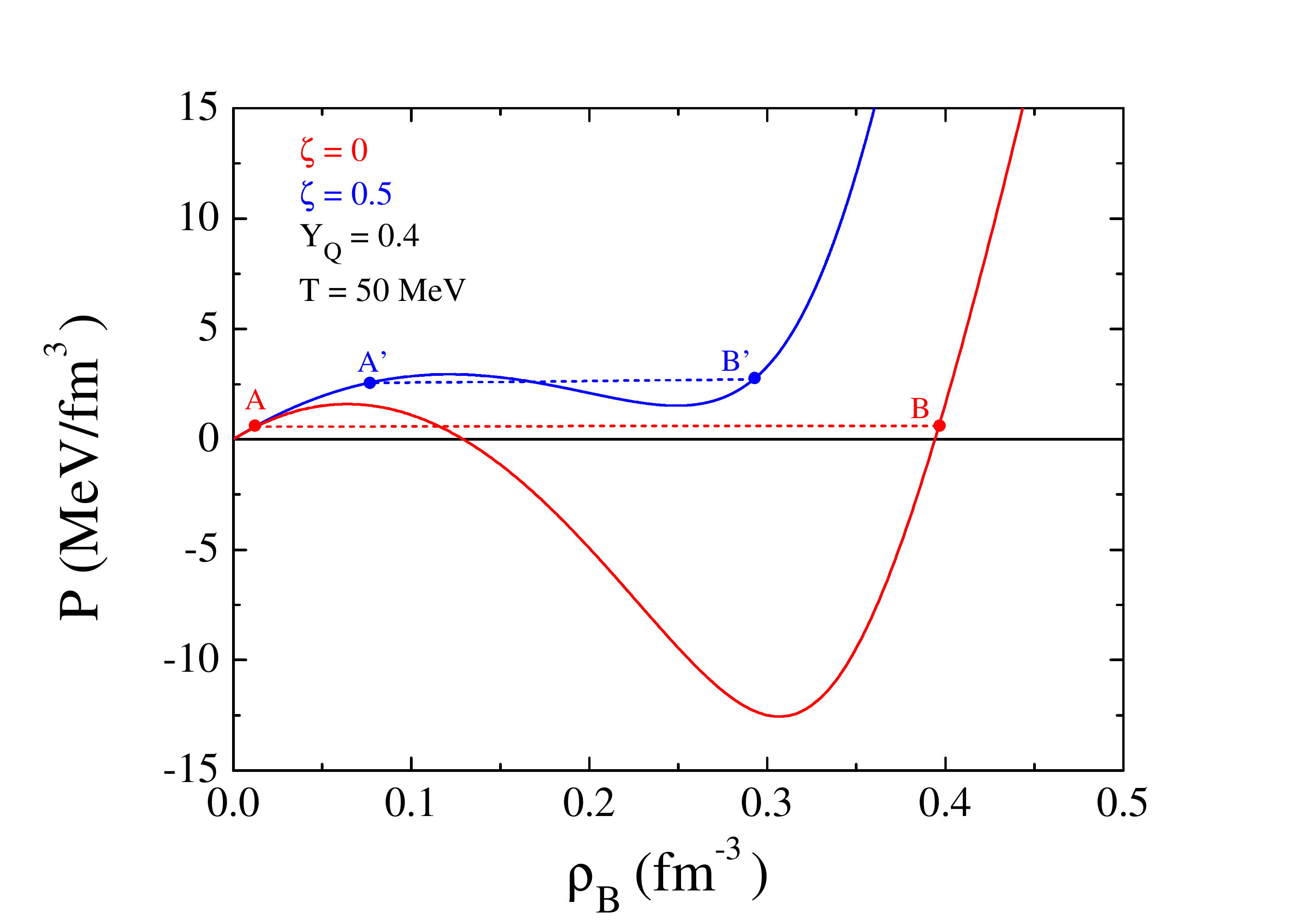}\hspace{-0.9cm}
	\includegraphics[width=0.49\linewidth]{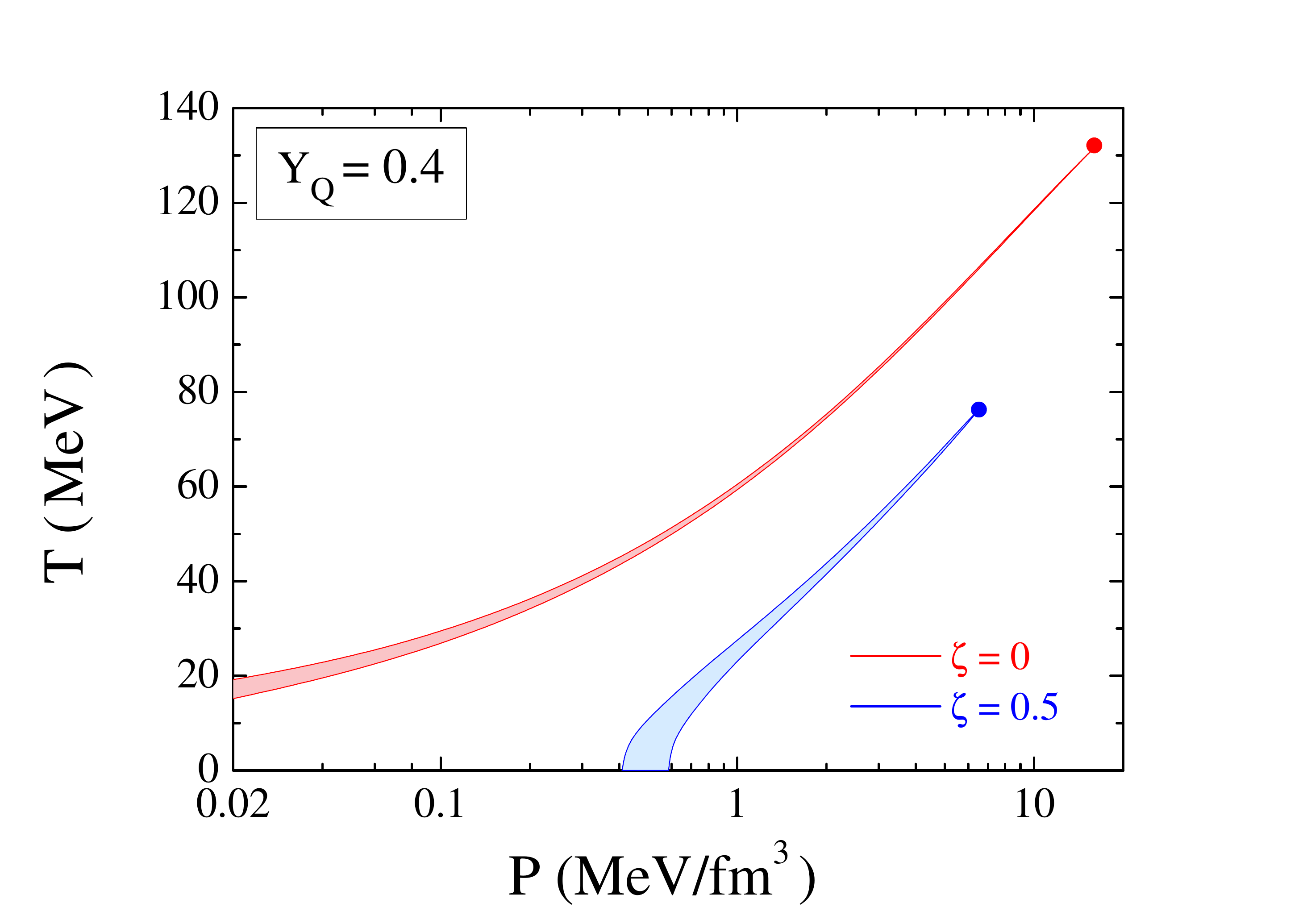}
	\caption{Left: phase transition for asymmetric matter: pressure versus 
					the baryonic density 	$\rho_B$ for $Y_Q=0.4$, $T=50$ MeV, and 
					$\zeta=G_V/G_S=0$ and $\zeta=G_V/G_S=0.5$. 
					The phase transition is indicated with a thin dashed line. 
          Since it is a two-component system the transition does not occur at 
          constant pressure. 
					The coordinates $(\rho_B,P)$ in (fm$^{-3}$, MeV fm$^{-3}$) of the 
					different points shown in the plot are $A(0.013,0.565),\,
					B(0.397,0.613)$, and $A'(0.076,2.56), \, B'(0.292,2.73)$.
					Right: transition lines  in the $T-P$ plane for $Y_Q=0.4$, with 
					$\zeta=0$ and 0.5.
          }
	\label{fig:02}
\end{figure*}

In the left panel of Fig. \ref{fig:02}, we plot the pressure as a function of the 
baryon density to exemplify a  phase transition from the broken phase to the
symmetric one at $T=50$ MeV, for $\zeta=0$ (curve $AB$, in red) and $\zeta=0.5$ (curve $A'B'$, in blue). This phase transition is obtained by keeping $Y_Q$ fixed during the transition; as a consequence, the pressure rises slightly. 
It also shows that the phase transition is weaker, i.e., corresponds to a smaller 
jump in density, when taking $\zeta=0.5$. 
This result is expected since the addition of the vector interaction is known to 
drag the first-order phase transition towards lower temperatures and make the 
pressure stiffer at smaller densities. For $\zeta>0.5$ and neutral matter with 
$Y_Q=0$, the first-order phase transition does not exist any more.

The right panel of Fig. \ref{fig:02} shows the phase diagram in the 
$T - P$ plane for both $\zeta=0$ and $\zeta=0.5$, with the charge ratio 
$Y_Q=0.4$. 
As discussed before, the pressure is not constant in a phase transition with a 
charge fraction different from 0 or 0.5.
The bands define the range of pressures covered during the phase transition.
When a finite repulsive vector interaction is considered the pressure at the 
transition increases considerably.

\begin{figure*}[t!]
	\centering
	\includegraphics[width=0.49\linewidth]{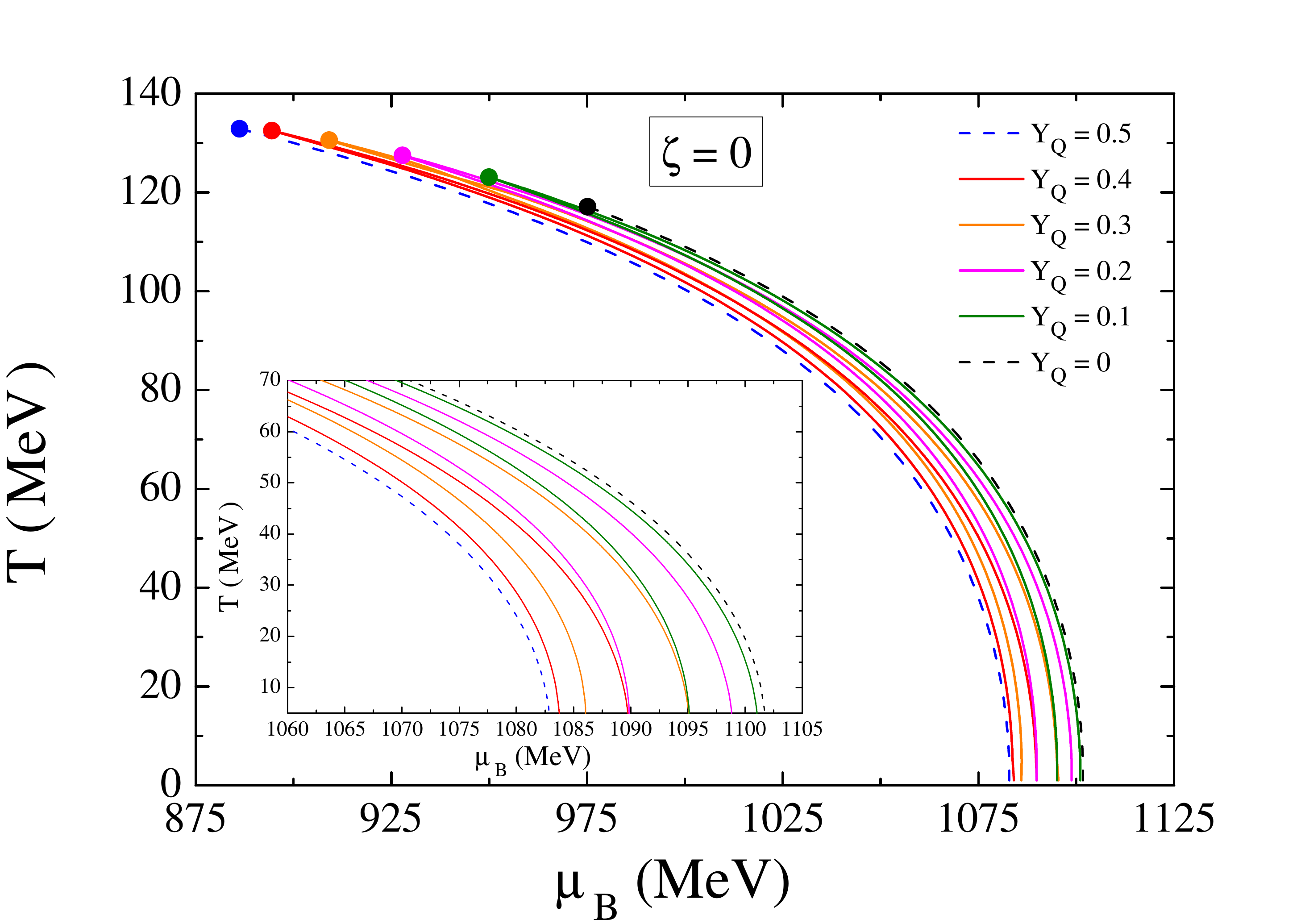}\hspace{-0.9cm}
	\includegraphics[width=0.49\linewidth]{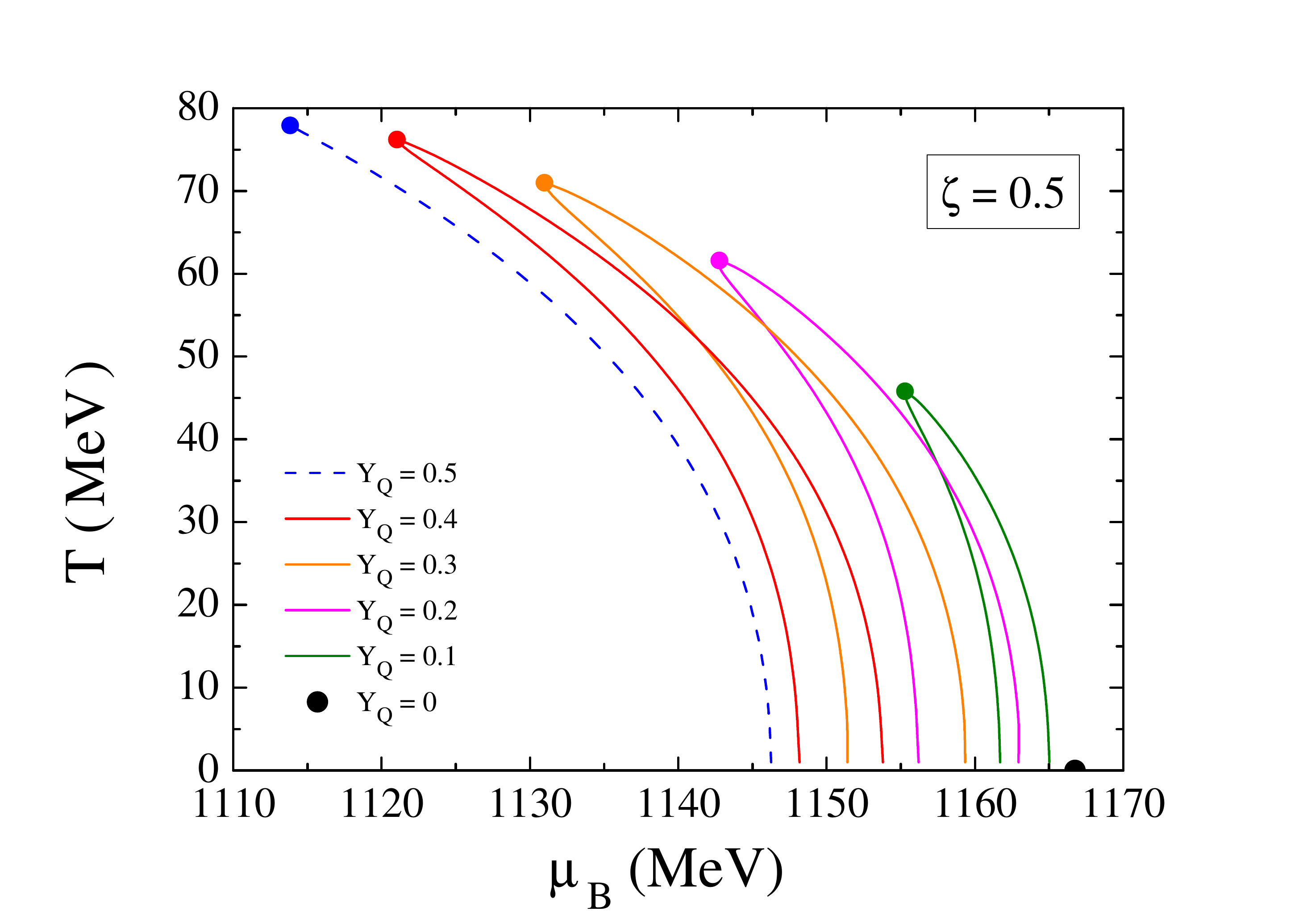}
	\includegraphics[width=0.49\linewidth]{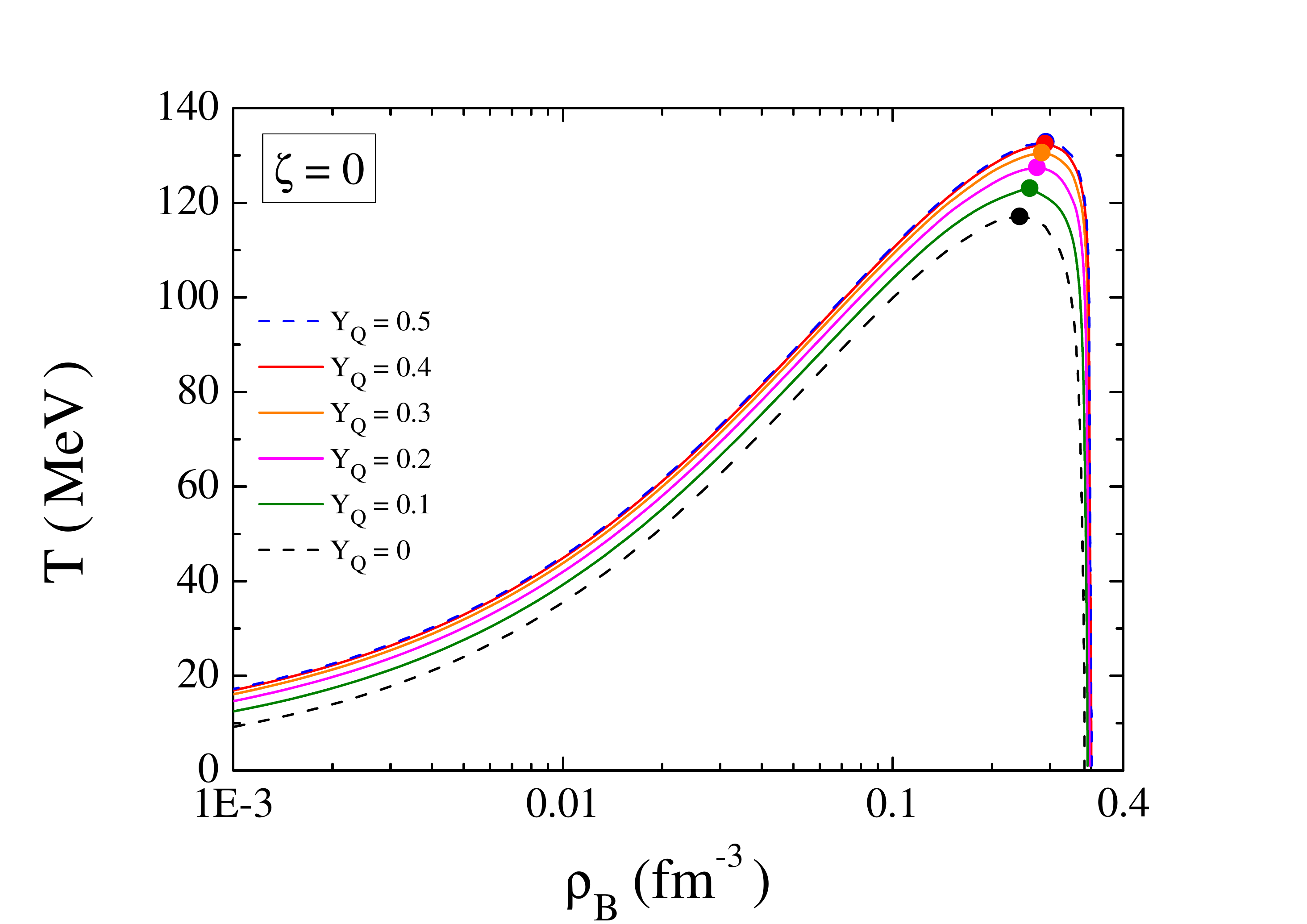}\hspace{-0.9cm}
	\includegraphics[width=0.49\linewidth]{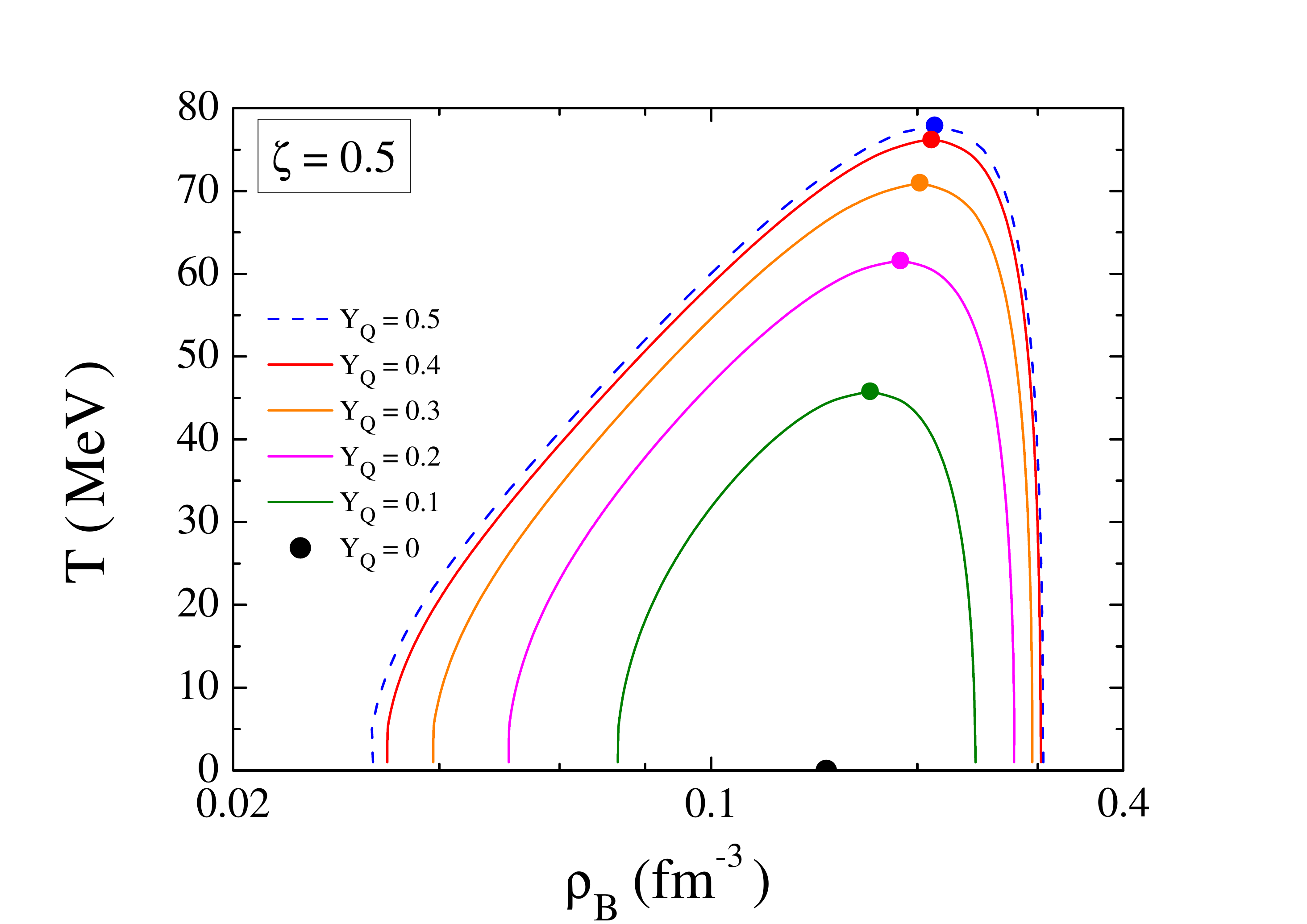}
	\caption{Transition lines in the $T-\mu_B$ plane (top panels) and
          $T-\rho_B$ plane (bottom panels) for $\zeta=0$ (left panels) and 
					$\zeta=0.5$ (right panels). The dots are the respective CEPs.
					In the two bottom panels a log scale is used on the $x$-axis. 
          }
	\label{fig:03}
\end{figure*}

In Fig. \ref{fig:03} the phase transition lines at different charge fractions 
are plotted for $\zeta=0$ (left panels) and $\zeta=0.5$ (right panels) in the 
$T-\mu_{B}$ plane (upper panels) and $T-\rho_{B}$ plane (lower panels). 
The two extremes $Y_Q=0$ and $Y_Q=0.5$ describe one-component matter: 
neutral matter with zero total electric charge and symmetric matter with 
equal amount of $u$ and $d$ quarks, respectively. In those extreme cases, 
identified in Fig. \ref{fig:03} by dashed lines, the phase transition occurs at
constant pressure and baryonic chemical potential and also reduces to a line in 
the $T-\mu_B$ plane.

For any other given $Y_Q$ in the $T-\mu_B$ or $T-\rho_B$ plane, the line at 
which matter starts to recover chiral symmetry and the line for which matter is 
already in a {\it partial} chiral-symmetric state\footnote{A partial chiral-symmetric state in the sense that the constituent quark mass is still far from 
the respective  current one.} are not coincident (see Fig. \ref{fig:03}). 
However, in the $T-\mu_{BQ}$ plane the phase transition is defined by a single line.

The effect of including a finite vector interaction is clear: for a 
given $Y_Q$, a finite $\zeta$ pushes the critical region and CEP toward higher 
values of  the chemical potential $\mu_{B}$ and lower values of the temperature. 
The main effect  on the phase transition of a charge fraction smaller
than 0.5 is seen in the localization of the CEP and the onset of the 
transition. In asymmetric matter,  i.e., with a smaller total charge, the CEP
moves to smaller temperatures and larger $\mu_{B}$ and the onset of
the transition moves to larger values of  $\mu_{B}$.  
For $\zeta=0$, the CEP chemical potential ($\mu_{B}^{\mathrm{CEP}}$) suffers a
rather large  change with the reduction of the total charge $Y_Q$: 
going from  $Y_Q=0.5$ to $Y_Q=0$, $\mu_{B}^{\mathrm{CEP}}$ varies $\approx 89$ MeV, 
while the temperature $T^{\mathrm{CEP}}$ is only moderately affected, suffering a 
reduction of $\approx16$ MeV. 
If $\zeta =0.5$, the CEP temperature experiences the largest variation  with
decreasing $Y_Q$, with a reduction of $\approx78$ MeV, while the change in 
$\mu_{BQ}^{\mathrm{CEP}}$  is not larger than 54 MeV. 
In this case, the CEP occurs at zero temperature  for $Y_Q=0$. 

There is no first-order phase transition  for $\zeta=0.5$ with $Y_Q=0$ because 
the CEP occurs at $T=0$. This is similar to neutron matter, for which there is 
no liquid-gas phase transition \cite{Muller:1995ji}. The  possible existence of 
a first-order  phase transition to a chirally symmetric state in neutral matter 
depends strongly on the coupling constant $G_V$, which plays a role similar to 
the one played by the $\rho$-meson coupling for nuclear relativistic 
mean-field models.

For all other charge fractions represented, the phase transition
is defined by two lines in the $T-\mu_B$ plane; see both top panels of
Fig. \ref{fig:03}. The range of temperatures and chemical potentials spanned by 
the $\zeta=0$ model is much larger, but the overall features are similar.
For each charge fraction, in the region between the left branch (at the lower 
chemical potential or density limit) and the right branch (at the larger 
chemical potential or density limit) matter separates into two phases: a low-density phase with broken chiral symmetry and a high-density   ({\it partially}) 
chiral-symmetric matter. 

In the bottom panels of Fig. \ref{fig:03}, the transition lines are represented 
in the $T-\rho_B$ plane and the presence of the vector interaction has a 
noticeable effect. The line that defines the left border of the transition 
region occurs at very low densities if $\zeta=0$ and $T\lesssim20$ MeV. On the 
other extreme, all lines come close to $\rho_B=0.4$fm$^{-3}$, which lies below 
3 times the saturation density.
The smaller the charge fraction, the lower the critical
temperature, the smaller the transition region and the smaller the
density at which chiral-symmetric matter sets in, although the differences 
are not  very large if $\zeta=0$. A different situation occurs for
$\zeta=0.5$; in this case, both the critical temperature and the
width of the transition decrease drastically with decreasing $Y_Q$,
and at $Y_Q=0$ the phase transition is reduced to a point (the CEP). 

An important conclusion is that for a finite  $\zeta$ the transition
to chiral-symmetric matter may occur at quite low densities, making
this phase more accessible experimentally. For matter characterized by
$Y_Q=0.4$ this is possible already at twice saturation density for a
small temperature, and at an even smaller density for larger
temperatures.  If during a nonequilibrium reaction  the system enters the 
region inside the binodal or spinodal, the density for the appearance of  chiral
symmetric matter will be even lower as we will discuss in the
following. In this case, however, in the form of clusterized matter.
For $\zeta=0$ the unstable/metastable region extends until very low densities, 
while for  $\zeta=0.5$ the low-density limit of the mixed phase  sets in
between 0.01 and 01 fm$^{-3}$  depending on the charge.

\begin{figure*}[t!]
  \centering
	\includegraphics[width=0.49\linewidth]{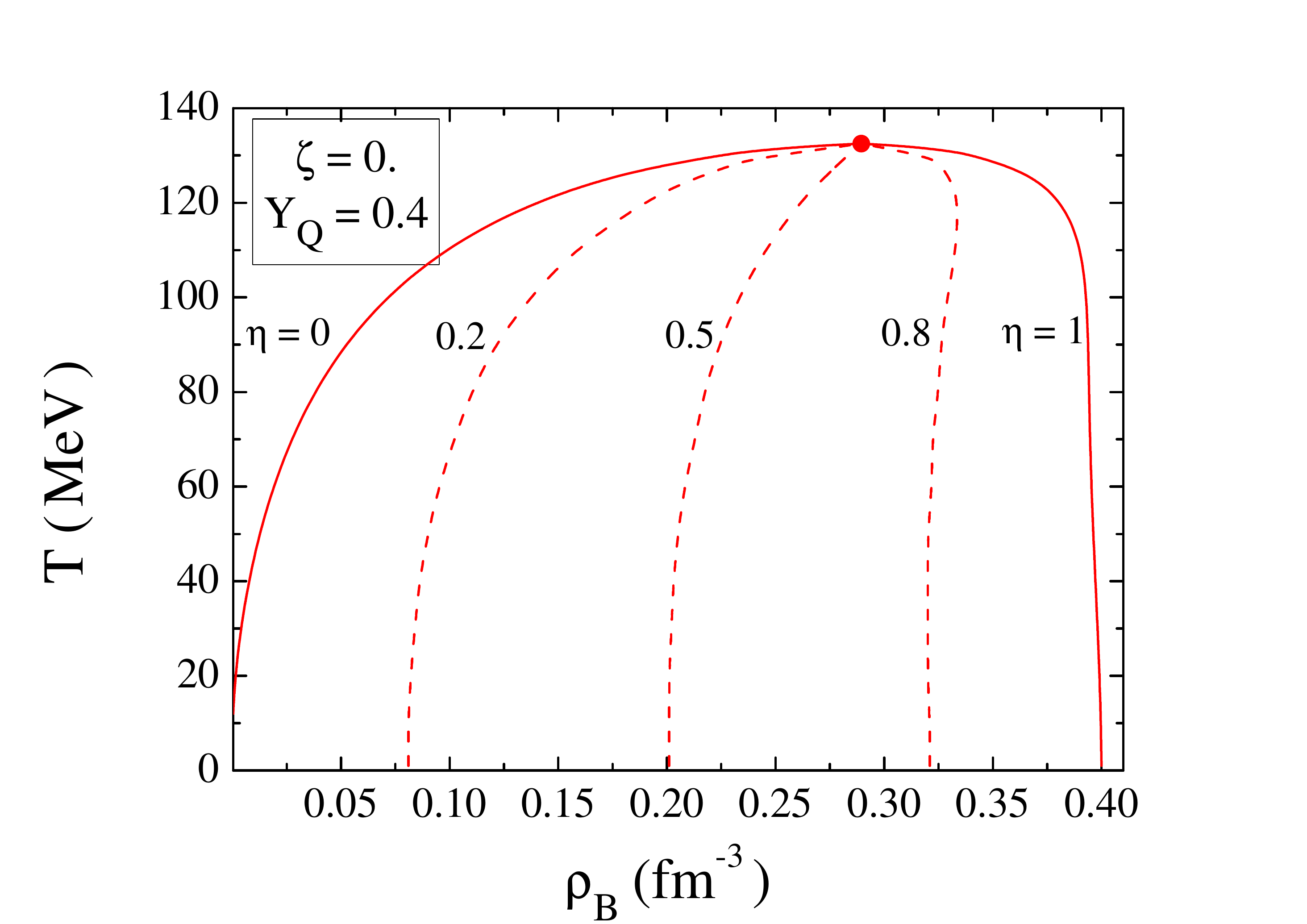}\hspace{-0.9cm}
	\includegraphics[width=0.49\linewidth]{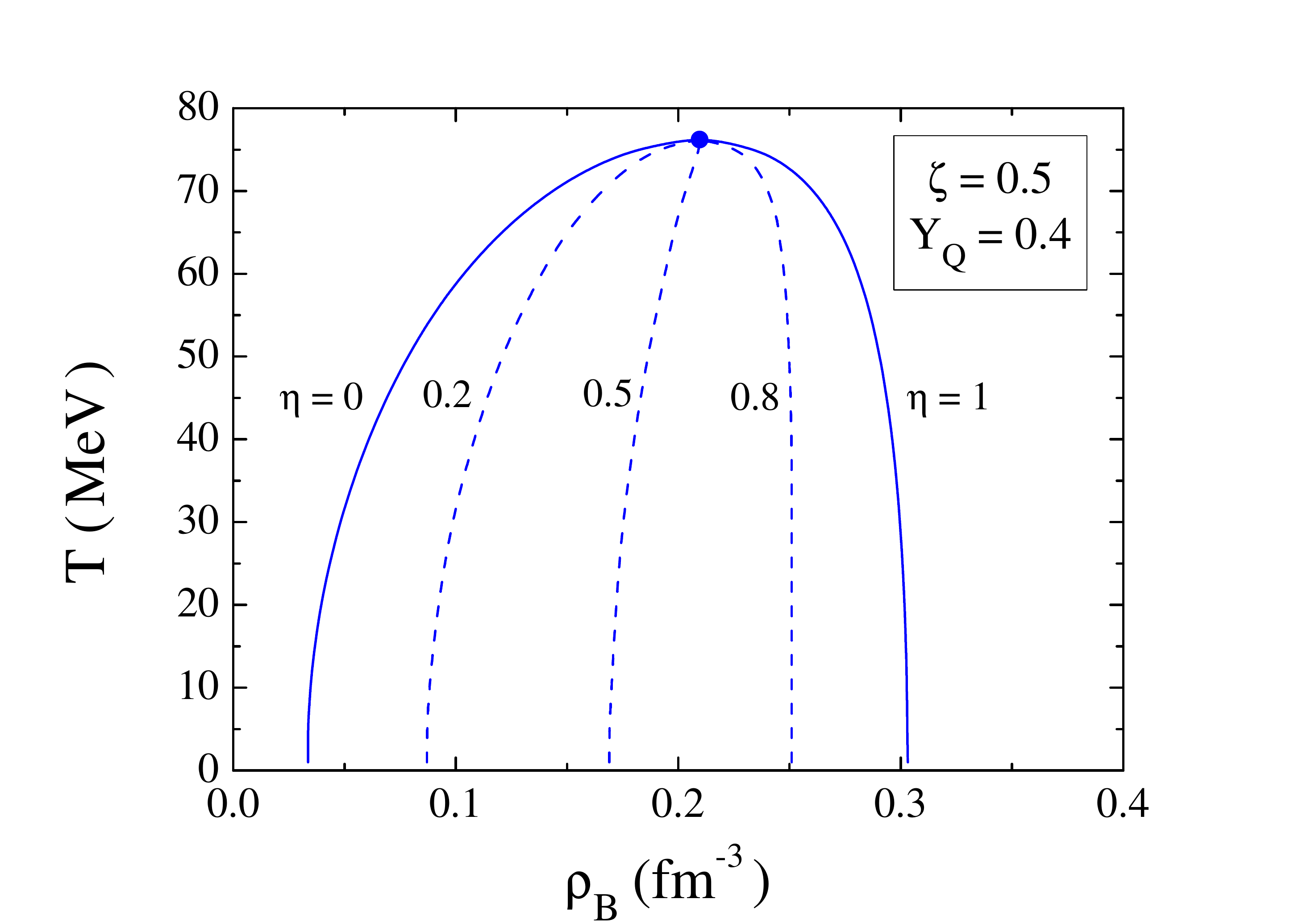}
	\caption{The transition lines (full lines) in the $T-\rho_B$ plane  
					for $\zeta=0$ (left) and $\zeta=0.5$ (right): 
					the left curve corresponds to matter in a chiral symmetric broken state 
          and the right curve to matter in a chiral-symmetric state. 
          The dashed lines identify the localization matter where 
          20\%, 50\%, and 80\% of matter is in chiral symmetric state, respectively. 
          } 
	\label{fig:04}
\end{figure*}

In Fig. \ref{fig:04} the phase diagrams in the $T-\rho_B$ plane are shown for 
$Y_Q=0.4$, with $\zeta=0$ (left panel) and $\zeta=0.5$ (right panel). 
The full lines identify the limits of the mixed phase: $\eta=0$ means that the 
matter is all in broken-chiral-symmetry phase, while matter with 
$\eta=1$ is in a completely restored-chiral-symmetry ({\it partially}) phase. 
Between the limiting lines of the mixed phase, several dashed lines
show where matter is constituted by different fractions of chiral-symmetric matter, identified by $\eta$. Within
the models we  have considered,  it is possible to form chiral-symmetric matter in clusterized matter at densities well below
the onset of pure chiral-symmetric matter, during a nonequilibrium evolution 
of the system.

\section{Conclusions}\label{sec:conclusions}

In this work we studied how the restoration of chiral symmetry, within
the PNJL model with and without vector interactions, is affected by
the conservation of more then one charge.  Although a three-flavor
quark model was considered, the critical region studied is only defined
by the up and down quarks, because the effect of the $s$ quarks is
only felt at higher energies than the ones considered.

During a  phase transition  involving  more than one conserved charge,
pressure is not  constant, indicating the presence of a mixed phase.
In this case, the transition region defined   in the  $T-\mu_B$  or
the  $T-\rho_B$ plane is limited by two distinct lines,
divided by a mixed phase. The width of this mixed phase is dependent
on the total charge fraction $Y_Q$ and heavily dependent on the
vector coupling $\zeta=G_V/G_S$.

The inclusion of a  vector interaction weakens the phase transition
both when considering a one-fluid system (such as those consisting of
symmetric matter or neutral matter) or a system with more than one
conserved charge (such as asymmetric quark matter with a charge fraction
$0<Y_Q<0.5$).  In particular, when taking a sufficiently strong vector coupling the
CEP disappears. 

It was shown that for asymmetric matter the restoration of chiral symmetry 
occurs at smaller densities, as suggested in Ref. \cite{Kaiser:2008qu}, and that the 
binodal section is reduced; in particular, the low-density onset and the high-density limits of the binodal move to larger and smaller densities, respectively. 
This effect is stronger if the vector interaction is included. 
Besides, the localization of the CEP also moves to lower temperatures.

During a nonequilibrium evolution, the system may reach regions of the phase 
diagram, such as the metastable or unstable regions inside the binodal and spinodal 
sections, that are forbidden to equilibrium thermodynamics \cite{Sasaki:2007db}. 
We have shown that under these conditions it is possible to form clusters made 
of chiral-symmetric matter at rather low densities. 
It is clear that these densities are model dependent. 
However, the fact that asymmetric matter favors the formation of chiral-symmetric matter at lower densities than the predictions from symmetric matter, 
is quite general.

These results are, in particular, of interest for neutron stars where the proton fraction is especially low. Besides, it was also shown that the occurrence of 
a quark phase inside  two solar-mass neutron stars requires the inclusion of the 
vector interaction in the Lagrangian density of NJL-like models; see, for 
instance, Refs. \cite{Bonanno:2011ch,Benic2014,Pereira:2016dfg}. 
These two characteristics--the asymmetry of matter and the presence of the 
vector interaction--may indicate that in hot environments (such as occurring in 
neutron star mergers or proto-neutron stars) the presence of chiral-symmetric 
quark matter is possible even at moderate densities. 
Carrying these conclusions to the laboratory, they seem to show that using large neutron-rich nuclei in heavy collisions  may give rise to conditions favorable 
to the formation of clusters of chiral-symmetric quark matter at intermediate 
densities attainable, for instance,  at FAIR or NICA. Even stable heavy nuclei 
such as lead or uranium with a charge fraction $\sim 0.4$ would already be 
sufficiently asymmetric to create the necessary conditions for quark matter 
formation. Unstable radioactive beams of very asymmetric heavy nuclei would 
further improve these conditions.

In the present study we have paid special attention to conditions presently 
attained at the RHIC, in particular, a charge fraction of 0.4 (as in Pb-Pb collisions) and zero strangeness fraction, i.e., the strangeness density is put to zero. 
In order to connect our results with lattice QCD calculations,
we have checked the leading-order contribution
in $\mu_B$ for the expansions of the electric charge and strangeness chemical potentials, $\mu_Q$ and $\mu_S$ (see Refs. \cite{Bazavov:2012vg,Borsanyi:2013hza}). For chemical potentials below 400 MeV and temperatures of 
the order of 200 MeV, these temperature-dependent coefficients are $q_1=-0.04$ and $s_1=0.34$. A more extended comparison with lattice QCD data, within a larger region of temperature and chemical potentials, is beyond the scope of this work and is left for upcoming investigations.

As future work, it would be interesting to study this scenario in a QCD model 
beyond the mean-field approximation. One way to incorporate quantum fluctuations 
in this calculation is by using the functional renormalization group (FRG). 
Recently, the QCD phase has been studied by applying the FRG approach to the quark 
meson model 
\cite{Schaefer:2004en,Herbst:2013ail,Tripolt:2013jra,Tripolt:2017zgc,CamaraPereira:2020xla}. The influence of strangeness neutrality on thermodynamic quantities such as the 
equation of state was studied in Ref. \cite{Fu:2018qsk}. 
Imposing strangeness conservation and the initial isospin asymmetry of the 
colliding nuclei is the natural step forward to continue these studies.

\section*{Acknowledgements}

This work was supported by national funds from FCT (Fundação para a Ciência e a 
Tecnologia, I.P, Portugal) under the IDPASC Ph.D. program (International 
Doctorate Network in Particle Physics, Astrophysics and Cosmology), with the 
Grant No. PD/BD/128234/2016 (R.C.P.), and under the Projects 
No. UID/FIS/04564/2019, No. UID/04564/2020, and No. POCI-01-0145-FEDER-029912 
with financial support from POCI “Programa Operacional Competitividade e 
Internacionalização (COMPETE 2020)”, in its FEDER component. 
Partial support comes from ``THOR'' (COST Action CA15213) and ``PHAROS'' 
(COST Action CA16214).








\bibliographystyle{apsrev4-1}
\bibliography{bib}
\end{document}